\documentclass[twocolumn,aps,superscriptaddress,amsmath,amssymb]{revtex4}
\usepackage{amsfonts}
\usepackage{bbm}
\usepackage{subfigure}
\usepackage{tipa}

\usepackage{graphicx}
\usepackage{dcolumn}
\usepackage{bm}

\begin{document}
\title{Spin-dependent electron transport along hairpin-like DNA molecules}
\author{Pei-Jia Hu}
\affiliation{Hunan Key Laboratory for Super-microstructure and Ultrafast Process, School of Physics and Electronics, Central South University, Changsha 410083, China}
\author{Si-Xian Wang}
\affiliation{Hunan Key Laboratory for Super-microstructure and Ultrafast Process, School of Physics and Electronics, Central South University, Changsha 410083, China}
\author{Xiao-Hui Gao}
\affiliation{Hunan Key Laboratory for Super-microstructure and Ultrafast Process, School of Physics and Electronics, Central South University, Changsha 410083, China}
\author{Yan-Yang Zhang}
\affiliation{School of Physics and Electronic Engineering, Guangzhou University, Guangzhou 510006, China}
\author{Tie-Feng Fang}
\affiliation{School of Physical Science and Technology, Lanzhou University, Lanzhou 730000, China}
\author{Ai-Min Guo}
\email[]{aimin.guo@csu.edu.cn}
\affiliation{Hunan Key Laboratory for Super-microstructure and Ultrafast Process, School of Physics and Electronics, Central South University, Changsha 410083, China}
\author{Qing-Feng Sun}
\affiliation{International Center for Quantum Materials, School of Physics, Peking University, Beijing 100871, China}
\affiliation{Collaborative Innovation Center of Quantum Matter, Beijing 100871, China}
\affiliation{CAS Center for Excellence in Topological Quantum Computation, University of Chinese Academy of Sciences, Beijing 100190, China}
\date{\today}
	
\begin{abstract}
The chirality-induced spin selectivity (CISS), demonstrated in diverse chiral molecules by numerous experimental and theoretical groups, has been attracting extensive and ongoing interest in recent years. As the secondary structure of DNA, the charge transfer along DNA hairpins has been widely studied for more than two decades, finding that DNA hairpins exhibit spin-related effects as reported in recent experiments. Here, we propose a setup to demonstrate directly the CISS effect in DNA hairpins contacted by two nonmagnetic leads at both ends of the stem. Our results indicate that DNA hairpins present pronounced CISS effect and the spin polarization could be enhanced by using conducting molecules as the loop. In particular, DNA hairpins show several intriguing features, which are different from other chiral molecules. First, the local spin currents can flow circularly and assemble into a number of vortex clusters when the electron energy locates in the left/right electronic band of the stem. The chirality of vortex clusters in each band is the same and will be reversed by switching the electron energy from the left band to the right one, inducing the sign reversal of the spin polarization. Interestingly, the local spin currents can be greater than the corresponding spin component of the source-drain current. Second, both the conductance and the spin polarization can increase with molecular length as well as dephasing strength, contrary to the physical intuition that the transmission ability of molecular wires should be poorer when suffering from stronger scattering. Third, we unveil the optimal contact configuration of efficient electron transport and that of the CISS effect, which are distinct from each other and can be controlled by dephasing strength. The underlying physical mechanism is illustrated.
\end{abstract}
	\maketitle

\section{\label{sec1}Introduction}

Recent works have made an important breakthrough, the so-called chirality-induced spin selectivity (CISS), in the emerging field of spintronics (see Refs.~\cite{nr1,ajm1,nr2} for a review). Since the original experiment by G\"{o}hler {\it et al.} that unpolarized photoelectrons passing through self-assembled monolayers of double-stranded DNA (dsDNA) are highly spin polarized at room temperature \cite{gb1}, the CISS effect has been validated by numerous established groups using different experimental techniques \cite{xz1,md1,dob1,ztj1,kv1,ajm2,dob2,akm1,km1,ms1,ajm3,sm1}. For instance, Xie {\it et al.} have reported by means of atomic force microscopy that a two-terminal dsDNA device behaves as an efficient spin filter, by performing direct charge transport through a single dsDNA sandwiched between two leads \cite{xz1}. Zwang {\it et al.} have designed electrochemical experiments to explore DNA-mediated charge transport, finding that the spin selectivity exhibits a diode-like switch when the handedness of dsDNA flips between right-handed $B$-form and left-handed $Z$-form \cite{ztj1}, as predicted by previous theoretical works \cite{gam1,gam2}. Using fluorescence microscopy, Abendroth {\it et al.} have investigated spin-dependent charge transfer in dsDNA assemblies bound with dye molecules, revealing that the fluorescence intensities are sensitive to the magnetization direction of underneath ferromagnetic substrates \cite{ajm2}. All these experiments reach a consensus of the CISS effect in dsDNA molecules, implying the equivalence of different experimental techniques. Motivated by these experimental observations, a number of theoretical models have been put forward to understand the CISS effect \cite{gam1,gam2,me1,eaa1,gam3,vs1,whn1,mas1,ay1,de1,mvv1,mk1,ds1,nd1,gm1,yx1,dgf1}. Although these models may differ from one to another, a general perspective emerges that the helical structure and the resulting spin-orbit coupling (SOC) are key factors to yield spin-selective electron transmission through chiral molecules.

\begin{figure}
\includegraphics[width=8.3cm]{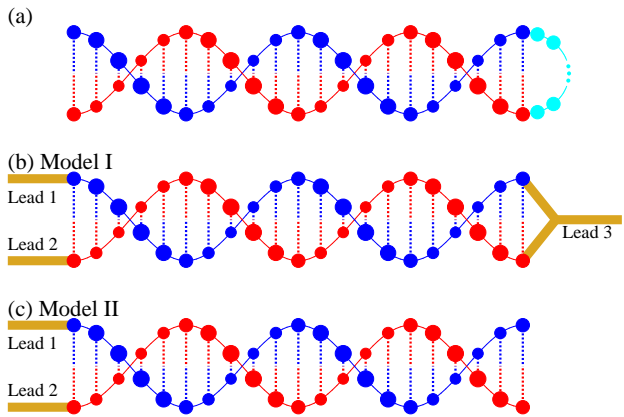}
\caption{\label{fig1} (a) Schematic of a DNA hairpin containing a Watson-Crick-paired stem (the double helix assembled from the blue and red balls) and an unpaired loop (the arc-shaped structure assembled from the cyan balls). All the blue and red balls represent the nucleobases, where the blue ones form the first helical chain and the red ones the second helical chain. The cyan balls could be the nucleobases or other organic groups. The dotted lines stand for the hydrogen bonding within Watson-Crick base pairs. (b)-(c) Two models used to simulate the spin transport properties of DNA hairpins, where each end of the hairpin stem is contacted by a normal-metal lead (lead 1 and lead 2, i.e., the source and the drain). Here, model I and model II correspond to the case that the hairpin loop exhibits conducting and insulating behavior, respectively (see text).}
\end{figure}

Aside from the traditional double helix studied extensively, DNA can self-assemble into other structural conformations such as hairpin. DNA hairpins consist of a Watson-Crick-paired stem and an unpaired loop, as illustrated by the double helix assembled from the blue and red balls and by the arc-shaped structure from the cyan balls in Fig.~\ref{fig1}(a), respectively. By attaching a donor chromophore to one end of the hairpin stem, photoinduced charge transfer in DNA hairpins has been widely investigated by taking diverse acceptor chromophores as the hairpin loop \cite{lfd1,lfd2,kk1,lfd3}. Besides the DNA sequence and the distance between donor and acceptor \cite{lfd4,gfc1,gfc2,rn1}, Renaud {\it et al.} have found the loop also plays an important role in the charge transfer along DNA hairpins \cite{rn2}. They have demonstrated a new charge transfer mechanism, deep-hole transfer, in DNA hairpins by choosing a suitable chromophore as the loop, where the holes propagate through low-lying electronic states of nucleobases and the charge transfer rates are enhanced by two orders of magnitude. In particular, recent experiments have reported the spin effects of DNA hairpins. Wasielewski {\it et al.} have prepared spin-entangled radical pairs within synthetic DNA hairpins possessing donor and acceptor chromophores, and probed their spin dynamics using electron paramagnetic resonance spectroscopy \cite{zta1,cr1,ojh1,ojh2}. These DNA hairpins can serve as molecular spin switches and are regarded as a promising platform for applications in quantum information science. Very recently, Stemer {\it et al.} have applied ultraviolet photoelectron spectroscopy to measure surface charging in photoemission from monolayers of mercurated DNA hairpins on ferromagnetic substrates, unveiling that photoionization energy depends on substrate magnetization orientation as well as the molecular handedness \cite{sdm1}, a signature of the CISS effect in DNA hairpins. Nevertheless, direct evidence of the CISS effect has not yet been provided in DNA hairpins.

In this paper, we report on a thorough study of spin-dependent electron transmission through DNA hairpins, which are connected by two nonmagnetic leads at the two ends of the stem, by considering various model parameters, as illustrated in Figs.~\ref{fig1}(b) and \ref{fig1}(c). In fact, previous works have already designed such DNA hairpins to study their conformational fluctuations, where both ends of the stem are chemisorbed by a fluorophore and a quencher. They have shown that these DNA hairpins are very stable at room temperature \cite{bg1,gnl1,wmi1} and can be used as molecular beacons to discriminate DNA molecules at single-nucleobase resolution \cite{ts1,ts2,db1}. We find that DNA hairpins behave as an efficient spin filter, especially when conducting molecules are chosen as the loop. In particular, they present several intriguing phenomena which are distinct from other chiral molecules. When the electron energy locates in the two electronic bands of the stem, the local spin currents can flow circularly in space, giving rise to a number of vortex clusters where a big vortex contains several small vortices. The chirality of vortex clusters in each oscillation region (electronic band) is identical and can be reversed by tuning the electron energy from the left oscillation region to the right one. It is interesting that the local spin currents can be much greater than the spin component of the source-drain current. Furthermore, contrary to the physical intuition that the transmission ability of molecular wires should be poorer when experiencing stronger scattering, the conductance of DNA hairpins can increase with molecular length as well as dephasing strength, so does the spin polarization. Finally, the optimal contact conditions of efficient electron transport and of the spin-selectivity effect are obtained, showing that they are different from each other and depend strongly on dephasing strength.

The rest of the paper is organized as follows. Section~\ref{sec2} introduces the model Hamiltonian of DNA hairpins with different loops, and the Green's function to calculate the spin polarization and the local spin currents. In Sec.~\ref{sec3}, the numerical results and discussion are presented. Section~\ref{sec3A} shows the dephasing effect on the spin transport along DNA hairpins and the distribution of the local spin currents, Section~\ref{sec3B} displays the effect of the molecular length on the spin transport, Section~\ref{sec3C} studies the influence of the interchain coupling on the spin transport, and Section~\ref{sec3D} investigates the contact effect on the spin transport. Finally, the results are summarized in Sec.~\ref{sec4}.

\section{\label{sec2} Model and Method}

The spin transport along DNA hairpins can be simulated by the Hamiltonian $\mathcal{H} =\mathcal{H}_{\rm hs} +\mathcal{H}_{\rm hl} + \mathcal{H} _{\rm d} +\mathcal{H}_{\rm ec}$. Here, $\mathcal{H}_{\rm hs}$ and $\mathcal{H}_{\rm hl}$ describe the hairpin stem and loop, respectively, $\mathcal{H} _{\rm d}$ is the dephasing term, and $\mathcal{H}_{\rm ec}$ represents the real leads including the molecule-lead coupling. The hairpin stem is identical to traditional dsDNA molecules and can be described by a two-leg ladder model with the SOC term \cite{gam1}:
\begin{equation}
\begin{aligned}
\mathcal{H}_{\rm hs}=& \sum_{j=1}^2 \left\{ \sum_{n=1}^N \varepsilon_{j} c_{j n}^{\dagger} c_{j n}+ \sum_{n=1}^{N-1} [i t_{\rm so} c_{j n}^{\dagger} (\sigma_{n}^{(j)} + \sigma_{n+1}^{(j)}) c_{j n+1} \right.\\ &\left. + t_{j} c_{jn}^{\dagger} c_{j n+1} + \mathrm{H.c.} ] \right\} + \sum_{n=1}^N (\lambda c_{1 n}^{\dagger} c_{2 n}+\mathrm{H.c.}), \label{eq1}
\end{aligned}
\end{equation}
where $c_{j n}^{\dagger} = (c_{jn \uparrow} ^\dagger, c_{jn \downarrow} ^\dagger)$ is the creation operator at site $\{j, n\}$ of the hairpin stem whose length is $N$, with $j$ labeling the helical chain and $n$ the base-pair index. $\varepsilon _{j}$ is the on-site energy, $t_{\rm so}$ is the SOC parameter, and $t_{j}$ ($\lambda$) is the intrachain (interchain) coupling. The SOC term is $\sigma_{n+1} ^{(j)}= \sigma_z \cos \theta -(-1)^j [\sigma_x \sin(n\Delta\varphi) - \sigma_y \cos(n \Delta\varphi)] \sin\theta$, with $\sigma_{x,y,z}$ the Pauli matrices, $\theta$ the space angle between the helical chain and the $x$-$y$ plane normal to the helix axis ($z$ axis), and $\Delta\varphi$ the twist angle between neighboring base pairs \cite{gam1}. The results obtained from the above Hamiltonian, Eq.~(\ref{eq1}), have shown that dsDNA molecules exhibit the CISS effect, the spin polarization increases with the molecular length, and no spin polarization appears in the single-stranded DNA (ssDNA) \cite{gam1}, in good agreement with the experiment \cite{gb1}.

Although a variety of ssDNA molecules and organic groups have been used as the hairpin loops \cite{lfd3,lfd4,gfc1,gfc2,rn1,rn2,zta1,cr1,ojh1,ojh2,sdm1,bg1,gnl1,wmi1,ts1,ts2,db1}, they can be divided into two categories, i.e., the conducting loops and the insulating ones. When conductive organics are covalently linked to the hairpin stem as the loop, such as the acceptor chromophores, the electrons can propagate through the loop. In this case, the hairpin loop is regarded as a nonmagnetic lead, lead 3 in Fig.~\ref{fig1}(b), with the net current flowing through lead 3 being zero, and the corresponding Hamiltonian is written as:
\begin{equation}
\mathcal{H} _{\rm hl}= \sum_{k} \varepsilon_{lk} l_k^\dagger l_k + t_{\rm hl} [l_k ^\dagger (c_{1N}+ c_{2N})+(c_{1N} ^\dagger+ c_{2N} ^\dagger) l_k]. \label{eq2}
\end{equation}
Here, $l_{k}^{\dagger}=(l_{k \uparrow} ^{\dagger}, l_{k \downarrow} ^{\dagger})$ is the creation operator of mode $k$ in the hairpin loop and $t_{\rm hl}$ is the coupling between the stem and the loop. The spin transport along DNA hairpins can then be described by model I, as illustrated in Fig.~\ref{fig1}(b). When the hairpin loop is insulating, such as ssDNA molecules, the electrons cannot transport through the loop and the Hamiltonian is written as:
\begin{equation}
\mathcal{H} _{\rm hl}= 0. \label{eq3}
\end{equation}
Then, the spin transport properties are simulated by model II, as can be seen in Fig.~\ref{fig1}(c).

The third term, $\mathcal{H}_{\rm d}$, is the Hamiltonian of dephasing which takes place naturally during the electron transport process. Such dephasing can be induced by the electron-phonon scattering and the electron-electron interaction. Besides, the electrons may be scattered from the nuclear spins and the foreign impurities as well. Notice that previous works have clearly demonstrated the decoherence in DNA hairpins \cite{lfd3,gfc1,rn1}. Such inelastic scatterings lead to the phase memory loss of the electrons and can be simulated by connecting each nucleobase of the hairpin stem to a B\"{u}ttiker's virtual lead \cite{gam1}.

Finally, the last term, $\mathcal{H}_{\rm ec}$, represents the real leads and their couplings to the hairpin stem. The Hamiltonian of $\mathcal{H}_{\rm ec}$ is written as:
\begin{equation}
\mathcal{H}_{\rm ec} =\sum_{k,j=1,2} \left[\varepsilon_{jk} a_{jk} ^{ \dagger} a_{jk} + t_{\rm ec} \left( a_{jk} ^{\dagger} c_{j1}+ c_{j1} ^{\dagger} a_{jk} \right) \right], \label{eq4}
\end{equation}
where $a_{jk}^{\dagger}=(a_{jk \uparrow} ^{\dagger}, a_{jk \downarrow} ^{\dagger})$ is the creation operator of mode $k$ in the $j$th lead, and $t_{\rm ec}$ is the coupling between the hairpin stem and the real leads. We stress that the real leads are attached to the two ends of the hairpin stem, as illustrated in Figs.~\ref{fig1}(b) and \ref{fig1}(c).

By employing the Landauer-B\"{u}ttiker formula, the current in lead $p$ (real or virtual) with spin $\sigma= \uparrow, \downarrow$ is expressed as \cite{ds1t}:
\begin{equation}
I_{p \sigma}=\frac {e^2} h \sum_{q, \sigma'} T_{p \sigma, q \sigma'} (V_{q}-V_{p}), \label{eq5}
\end{equation}
where $V_p$ is the voltage of lead $p$ and
\begin{equation}
T_{p\sigma, q\sigma'}={\rm Tr} [\mathbf{\Gamma}_{p\sigma} \mathbf{G}^r \mathbf{\Gamma}_{q \sigma'} \mathbf{G}^a ] \label{eq6}
\end{equation}
is the transmission coefficient from lead $q$ with spin $\sigma'$ to lead $p$ with spin $\sigma$. The Green's function $\mathbf{G} ^r (E)=[\mathbf{G} ^a (E)]^\dagger=[E \mathbf{I}- \mathbf{H}_ {\rm hs}- \sum_{p \sigma} \mathbf{\Sigma}_{p \sigma}^r ]^{-1}$ and the linewidth function $\mathbf{\Gamma} _{p\sigma}=i [\mathbf{\Sigma}_ {p\sigma} ^r- (\mathbf{\Sigma} _{p\sigma}^r) ^\dagger]$, with $E$ the Fermi energy and $\mathbf{\Sigma}_{p \sigma}^r$ the retarded self-energy due to the coupling to lead $p$. Here, we consider the wide-band limit, where the retarded self-energy is set to $\Sigma_ {p\sigma} ^r =-i \Gamma/ 2$ for the real leads, $\Sigma_ {p\sigma} ^r =-i \Gamma_h/ 2$ for lead 3, and $\Sigma_ {p\sigma} ^r =-i \Gamma_d/ 2$ for virtual leads, with $\Gamma_{d}$ the dephasing strength. Under the boundary condition that the net current flowing through lead 3 and each virtual lead is zero, their voltages can be calculated from Eq.~(\ref{eq5}) by applying a small external bias $V_0$ between the source and the drain, with $V_1=V_0$ and $V_{2}=0$. Finally, the spin-up and spin-down conductances are written as:
\begin{equation}
G_\sigma=\frac {e^2} h \sum_{q, \sigma'} T_{2 \sigma, q \sigma'} \frac {V_q} {V_0}, \label{eq7}
\end{equation}
and the spin polarization is defined as:
\begin{equation}
P_s=\frac {G_{\uparrow}- G_{\downarrow} } {G_{\uparrow}+ G_{\downarrow} }. \label{eq8}
\end{equation}

Besides, the local spin current flowing from site $m$ with spin $\sigma$ to its neighboring one $n$ with $\sigma'$ is also calculated \cite{jap1,jh1}:
\begin{equation}
I_{m \sigma \rightarrow n \sigma'}= -\frac {2e} h \int_ {-\infty} ^{+\infty} {\rm Re} [ (\mathbf{H}_{\rm hs})_{m \sigma,n \sigma'} \mathbf{G} _{n \sigma', m\sigma}^< (\xi)] d \xi, \label{eq9}
\end{equation}
where $\mathbf{G} ^<$ is the Keldysh Green's function. By using the Keldysh equation $\mathbf{G} ^< = \mathbf{G} ^r (\sum_p i \mathbf{\Gamma}_p f_p) \mathbf{G}^a$, with $f_p$ the Fermi-Dirac distribution function and $\mathbf{\Gamma}_p= \sum_\sigma \mathbf{\Gamma}_ {p\sigma}$, Eq.~(\ref{eq9}) can be rewritten as:
\begin{eqnarray}
I_{m \sigma \rightarrow n \sigma'} &=& \frac {2e} h \sum_{p} \int_{-\infty} ^0 {\rm Im} \big[  (\mathbf{H}_{\rm hs})_{m \sigma, n \sigma'} ( \mathbf{G}^r  \mathbf{\Gamma} _p \mathbf{G}^a)_{n \sigma', m\sigma} \big] d\xi \nonumber \\ &+& \frac {2e^2} h \sum_{p } {\rm Im} \big[ (\mathbf{H}_{\rm hs}) _{m \sigma, n \sigma'} \mathbf{G}_{n \sigma', m\sigma} ^p(E) V_p \big], \label{eq10}
\end{eqnarray}
where $\mathbf{G}^p (E)= \mathbf{G}^r (E) \mathbf{\Gamma}_ p \mathbf{G} ^a (E)$ is the electron correlation function. The first term of Eq.~(\ref{eq10}) denotes the equilibrium persistent current $I^{eq}_{m \sigma \rightarrow n \sigma'}$ \cite{persistC1,persistC2} and the total equilibrium charge current $\sum_{\sigma,\sigma'} I^{eq}_{m \sigma \rightarrow n \sigma'}$ is equal to zero because the time-reversal symmetry is preserved for DNA hairpins. Then, the local transport spin current between neighboring sites $m$ with $\sigma$ and $n$ with $\sigma' $ can be simplified as:
\begin{equation}
I_{m \sigma \rightarrow n \sigma'}=\frac {2e^2} h {\rm Im} \sum_{p} \big[ (\mathbf{H}_{\rm hs})_{m\sigma, n\sigma'} (\mathbf{G}^r \mathbf{\Gamma}_ p \mathbf{G} ^a )_{n\sigma', m\sigma} V_p\big], \label{eq11}
\end{equation}
and the local spin-up and spin-down currents are defined as:
\begin{equation}
I_{m \rightarrow n} ^\sigma =\sum_{\sigma'} I_{m \sigma' \rightarrow n \sigma}. \label{eq12}
\end{equation}
The local current can be obtained by summing over the spin indices, which is consistent with previous works \cite{ns1,zyy1}. The total spin current flowing through DNA hairpins can be calculated by summing the local spin current over the connection between other sites and site $\{2,1\}$ which is contacted by the drain. Then, we can obtain the spin polarization, which is the same as Eq.~(\ref{eq8}), demonstrating the correctness of our numerical results.

\section{\label{sec3}RESULTS AND DISCUSSION}

For DNA hairpins, the model parameters are taken as $\varepsilon _{1}=0$, $\varepsilon _{2}=0.3$, $t_{\rm so}=0.01$, $t_{1 }=0.12$, $t_{2 }=-0.1$, $\lambda=-0.3$, with the unit eV, and $N=30$. The structural parameters are set to $\theta \approx 0.66$ and $\Delta \varphi=\pi / 5$. For real leads, the parameter $\Gamma=1$; for lead 3, $\Gamma_h=1$; for virtual leads, the dephasing strength $\Gamma_{d}=0.005$. All these parameters are the same as previous work \cite{gam1}, allowing direct comparison between spin transport properties of DNA hairpins and dsDNA molecules. The parameters $\lambda$, $N$, $\Gamma$, and $\Gamma_d$ will be used throughout the paper, unless stated otherwise. Our results indicate that DNA hairpins exhibit pronounced spin-filtering effect in a very wide range of model parameters and unique features which are different from dsDNA molecules.

\subsection{\label{sec3A}Dephasing effect on spin transport along DNA hairpins}

\begin{figure}
\includegraphics[width=8.3cm]{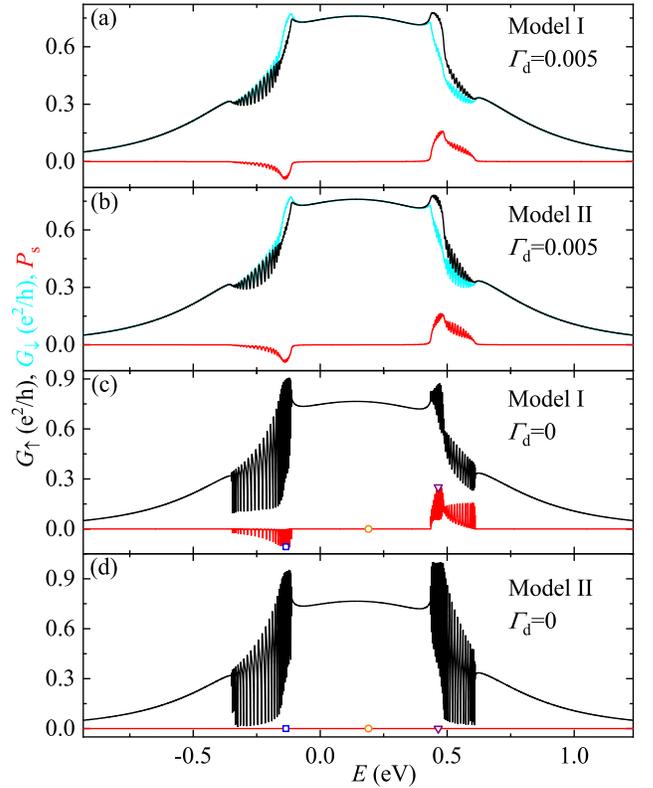}
\caption{\label{fig2} Spin transport along DNA hairpins in the presence and absence of B\"{u}ttiker's virtual leads. Energy-dependent spin-up conductance $G_\uparrow$ (black lines), spin-down conductance $G_\downarrow$ (cyan lines), and spin polarization $P_s$ (red lines) for (a) model I and (b) model II with $\Gamma_d=0.005$. $G_\uparrow$ vs $E$ (black lines) and $P_s$ vs $E$ (red lines) for (c) model I and (d) model II with $\Gamma_d=0$. Here, the symbols of squares, circles, and down triangles in (c) and (d) indicate electron energies at which the local spin current distributions are displayed in Fig.~\ref{fig3}.}
\end{figure}

We first consider the dephasing effect on the spin transport along DNA hairpins. Figures~\ref{fig2}(a) and~\ref{fig2}(b) show the spin-up conductance $G_\uparrow$ (black lines), spin-down conductance $G_\downarrow$ (cyan lines), and spin polarization $P_s$ (red lines) for model I and model II, respectively, as a function of the electron energy $E$ in the presence of dephasing. It clearly appears that the transmission spectra of DNA hairpins are completely different from the dsDNA molecules \cite{gam1}, although the model parameters of these two systems are identical. For both models of DNA hairpins, the conductances are nonzero over the entire energy spectrum which can be definitely classified into two categories, the oscillation regions at $[-0.35,-0.11] \cup [0.43,0.61]$ and the smooth ones in the remaining areas. We stress that these two oscillation regions coincide exactly with the electronic bands of the dsDNA molecules \cite{gam1}. One can see that both $G_\uparrow $ and $ G_\downarrow$ oscillate with $E$ in the oscillation regions, owing to the quantum interference effect. The oscillation frequency, i.e., the density of the transmission peaks, becomes higher when $E$ is close to the central smooth region, which is related to the SOC effect. Meanwhile, the conductances tend to decrease as the electron energy in the right (left) oscillation region is shifted towards higher (lower) energies, because of the increment of the scattering from the nucleobases.

The transmission profiles are, however, different in the smooth regions which are divided into three parts by the oscillation ones and absent in the dsDNA molecules \cite{gam1}. In the smooth regions where $E$ locates beyond the electronic bands, the oscillating behavior of $G_{\uparrow /\downarrow}$ vs $E$ vanishes as the quantum interference effect is negligible in these areas. In the central smooth region where $E$ is close to the potential energies of all the nucleobases, $G_{\uparrow /\downarrow}$ is very large and roughly independent of $E$. While in the other smooth regions where $E$ is far away from the potential energies, the electrons suffer strong scattering and $G_{\uparrow /\downarrow}$ becomes small. By further separating $E$ from the center of the energy spectrum, $G_{\uparrow /\downarrow}$ is declined monotonically as the scattering is gradually enhanced.

Apart from the unique features mentioned above, one can see other important phenomena. In the oscillation regions, $G_\uparrow $ is obviously distinct from $G_\downarrow $ for both models, indicating the spin-filtering effect of DNA hairpins. The spin polarization is nonzero over the whole oscillation regions and can achieve 15.9\% (16.3\%) for model I (II). In particular, $P_s$ oscillates with $E$ as well [see the bumps of the red lines in Figs.~\ref{fig2}(a) and~\ref{fig2}(b)], which cannot be observed in the dsDNA molecules \cite{gam1,gam3}. In the right (left) oscillation region, $P_s$ is positive (negative), which may correspond to the holes (electrons). In fact, a critical value $E_0\simeq 0.19$ is found in the curve $P_s$-$E$ that $P_s>0$ for $E>E_0$ and $P_s<0$ for $E<E_0$. When the holes migrate along the direction from, e.g., the source to the drain, it could be regarded as the situation of electron transmission along the opposite direction, leading to the sign reversal of $P_s$ in the two oscillation regions. Whereas in the smooth regions, no observable difference can be detected between $G_\uparrow$ and $G_\downarrow$, and $P_s$ is very small with the order of magnitude being $10^{-4}$. Except for these similarities, one can see that the oscillating amplitude of $G_{\uparrow /\downarrow} /P_s$ is smaller in model I as compared with model II, because of the extra lead---lead 3---in the former model. This is a universal phenomenon which holds for other model parameters, as demonstrated in the following numerical results, indicating that lead 3 in multi-terminal DNA devices can induce dephasing as well, just like the B\"{u}ttiker's virtual leads. Therefore, the electrons will lose their phase memory more quickly in model I.

\begin{figure*}
\includegraphics[width=17.8cm]{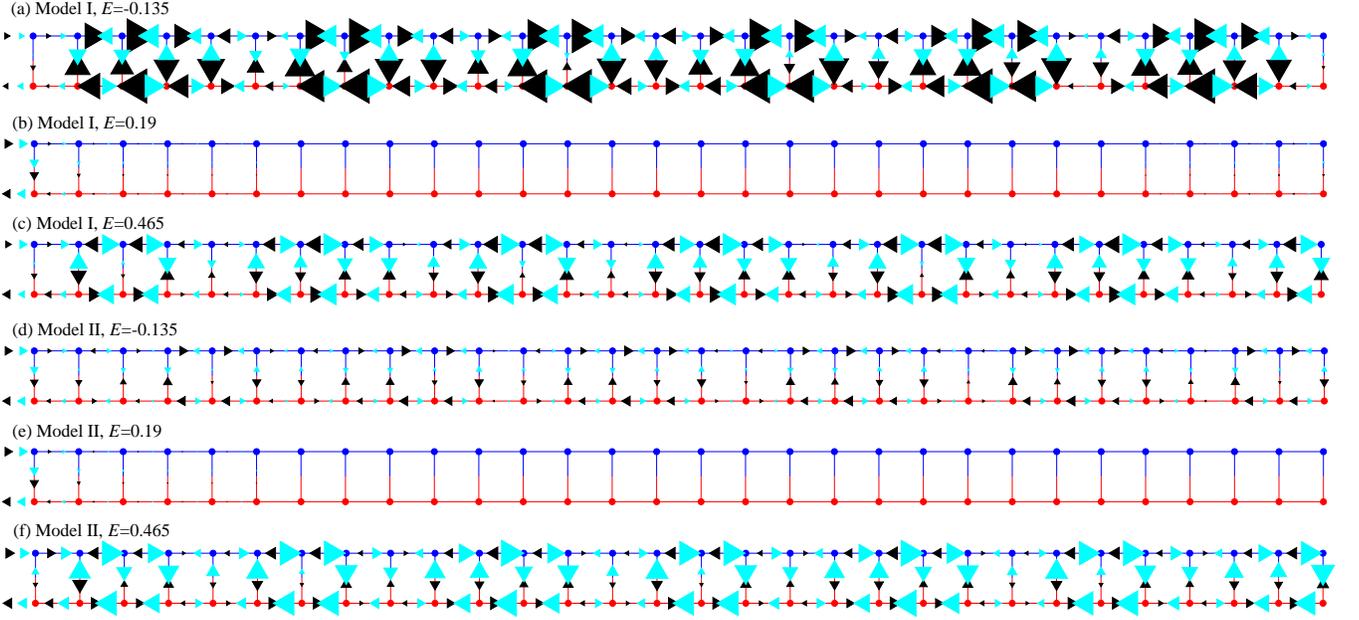}
\caption{\label{fig3} Spatial distributions of the local spin currents for DNA hairpins, in the case of $\Gamma _d=0$, at typical electron energies marked by the squares, circles, and down triangles in Figs.~\ref{fig2}(c) and \ref{fig2}(d). Spatial distributions of the local spin-up current $I_{m\rightarrow n}^ \uparrow$ (black arrows) and the local spin-down one $I_{m \rightarrow n}^ \downarrow$ (cyan arrows) for model I at (a) $E=-0.135$, (b) $E=0.19$, and (c) $E=0.465$. Spatial distributions of $I_{m\rightarrow n}^ \uparrow$ (black arrows) and $I_{m \rightarrow n}^ \downarrow$ (cyan arrows) for model II at (d) $E=-0.135$, (e) $E=0.19$, and (f) $E=0.465$. Here, the blue and red circles represent the nucleobases in the first and second helical chains, respectively, just the same as Fig.~\ref{fig1}. The size of the arrows is proportional to the magnitude of the local spin currents.}
\end{figure*}

To provide insight into the spin-filtering effect of DNA hairpins, Figs.~\ref{fig2}(c) and~\ref{fig2}(d) show $G_\uparrow$ and $P_s$ for model I and model II, respectively, as a function of $E$ by disconnecting to any B\"{u}ttiker's virtual lead, $\Gamma_d=0$. It is obvious that the transmission spectra of DNA hairpins can also be classified into the oscillation regions and the smooth ones. However, one can identify several distinct features in the absence of B\"{u}ttiker's virtual leads. The oscillating amplitude of $G_\uparrow$ is significantly enhanced [see the black line in Fig.~\ref{fig2}(c)], because the electrons do not experience any inelastic scattering from the nucleobases except for the last base pair and the quantum interference effect is strengthened. Interestingly, $P_s$ is also nonzero in the oscillation regions and can reach greater value of 25.8\% [see the bumps of the red line in Fig.~\ref{fig2}(c)], which is comparable to the dsDNA molecules \cite{gam1}. This is attributed to the fact that the device is switched into a multi-terminal one in the presence of lead 3 which plays a similar role as B\"{u}ttiker's virtual leads, giving rise to the spin-filtering effect of DNA hairpins. Contrarily, $P_s$ is declined to zero exactly in the smooth regions. While for model II, the oscillating amplitude of $G_\uparrow$ is even larger, as the electron transport process is completely coherent, due to the absence of lead 3 and B\"{u}ttiker's virtual ones. In this case, no spin polarization appears [see the red line in Fig.~\ref{fig2}(d)], regardless of the model parameters and consistent with previous works \cite{gam1,mas1,ay1}. This arises from the time-reversal symmetry and the phase-locking effect in two-terminal devices \cite{gam3,sqf1}.

In order to further understand the physical nature of the spin-filtering effect, the local spin currents of DNA hairpins with finite $P_s$ are calculated and compared with the situation of $P_s=0$. To this end, we consider both models in the absence of B\"{u}ttiker's virtual leads. Figures~\ref{fig3}(a)-\ref{fig3}(c) plot the spatial distributions of the local spin-up current $I _{m \rightarrow n} ^\uparrow $ (black arrows) and the local spin-down one $I _{m \rightarrow n} ^\downarrow $ (cyan arrows) for model I at typical electron energies which are marked by the square, circle, and down triangle in Fig.~\ref{fig2}(c), where $P_s\simeq-10.7\%$, 0, and 25.1\%, respectively. As a comparison, Figs.~\ref{fig3}(d)-\ref{fig3}(f) dsiplay the local spin current distributions for model II at the same electron energies with $P_s=0$. Here, the size of the arrows is proportional to the magnitude of the local spin currents. We find that the local current flowing from site $\{1, n\}$ to $\{1, n+1\}$ is always identical to that from $\{2, n+1\}$ to $\{2, n\}$, i.e.,
\begin{equation}
\sum_{\sigma= \uparrow, \downarrow} I_{\{1,n\} \rightarrow \{1, n+1\}} ^\sigma \equiv \sum_{\sigma= \uparrow, \downarrow} I_{\{2,n+1\} \rightarrow \{2, n\}} ^\sigma
\end{equation}
for $n \in [1,N-1]$, because of the current conservation. Besides, $G_{\uparrow /\downarrow}$ and $P_s$ obtained from Fig.~\ref{fig3} are the same as Fig.~\ref{fig2}, demonstrating the validity of our numerical results.

It is clear that in the oscillation regions, the electrons can propagate along the longitudinal direction of DNA hairpins [Figs.~\ref{fig3}(a), \ref{fig3}(c), \ref{fig3}(d), and \ref{fig3}(f)], although the source and drain are connected to their left side [Figs.~\ref{fig1}(b) and \ref{fig1}(c)], because $E$ locates within the electronic bands of the hairpin stem. The local spin-up and spin-down currents flow almost in opposite directions except for only a few nucleobases neighboring to the source/drain [see the first and second base pairs in Figs.~\ref{fig3}(a), \ref{fig3}(c), \ref{fig3}(d), and \ref{fig3}(f)]. Interestingly, the local spin currents can flow circularly in space, giving rise to vortex-like patterns, which have been reported in other molecular systems \cite{ns1,zyy1,sgc1}. Let us consider $I _{m \rightarrow n} ^\uparrow $ in Fig.~\ref{fig3}(a) as an example and similar phenomena can be observed for $I _{m \rightarrow n} ^\downarrow $. The local spin-up current can flow along the circular pathway
\begin{align*}
\{1,2\} & \rightarrow \{1,3\} \rightarrow \{1,4\} \rightarrow \{1,5\}  \rightarrow \{2,5\} \\ \nonumber
& \rightarrow \{2,4\} \rightarrow \{2,3\} \rightarrow \{2,2\} \rightarrow \{1,2\},
\end{align*}
and form a clockwise vortex [see the black arrows in the leftmost part of Fig.~\ref{fig3}(a)]. Inside such a large vortex, there exist three small vortices, such as
\begin{align*}
\{1,2\} & \rightarrow \{1,3\} \rightarrow \{1,4\} \rightarrow \{2,4\} \\ \nonumber
& \rightarrow \{2,3\} \rightarrow \{2,2\} \rightarrow \{1,2\}
\end{align*}
and
\begin{align*}
\{1,3\} \rightarrow \{1,4\} \rightarrow \{2,4\} \rightarrow \{2,3\} \rightarrow \{1,3\},
\end{align*}
all of which have the same chirality as the large vortex. Subsequently, a vortex cluster appears where a big vortex contains several small vortices. The vortex clusters in either oscillation region possess identical chirality and inside a single cluster different vortices will interfere with each other, leading to the instructive quantum interference effect. As a result, $G_\uparrow$ oscillates with $E$ in the electronic bands [see the black lines in Figs.~\ref{fig2}(c) and \ref{fig2}(d)], and the local spin currents inside a vortex cluster can be much greater than the corresponding spin component of the source-drain current [Figs.~\ref{fig3}(a), \ref{fig3}(c), and \ref{fig3}(f)]. Besides, two neighboring vortex clusters are usually separated by a single vortex with opposite chirality. For instance, the aforementioned vortex cluster is separated from its neighboring one by a counterclockwise vortex
\begin{equation*}
\{1,6\} \rightarrow \{1,5\} \rightarrow \{2,5\} \rightarrow \{2,6\} \rightarrow \{1,6\}.
\end{equation*}

By inspecting Figs.~\ref{fig3}(a)-\ref{fig3}(c),
it is important to notice that the local spin currents flowing
from $\{1, 29\}$ ($\{2, 29\}$) to $\{1, 30\}$ ($\{2, 30\}$)
are unequal to the corresponding spin current flowing within
the last base pair, which holds for all $E$'s.
In other words, a spin-flip-like process takes place at the last base pair
when it is connected to lead 3, which is caused by the spin phase memory loss
of the electrons when transmitting through this extra lead.
In contrast, such process cannot occur at the last base pair
in the absence of lead 3 [Figs.~\ref{fig3}(d)-\ref{fig3}(f)].
When $E$ is adjusted from the left oscillation region to the right one,
the chirality of the vortex clusters is reversed [Figs.~\ref{fig3}(a) and \ref{fig3}(c)],
which may correspond to the case that the type of charge carriers is switched
from the electrons to the holes, thus inducing the sign reversal of $P_s$.
It is interesting that in the left oscillation region where $G_\uparrow$ is smaller than $G_\downarrow$ with $P_s<0$, conversely $I _{m \rightarrow n} ^\uparrow $ is larger than the corresponding $I _{m \rightarrow n} ^\downarrow $ [Fig.~\ref{fig3}(a)]; while in the right oscillation region where $G_\uparrow$ is larger than $G_\downarrow$, $I _{m \rightarrow n} ^\uparrow $ is smaller than $I _{m \rightarrow n} ^\downarrow $ [Fig.~\ref{fig3}(c)].

Regarding the smooth regions, however, the local spin current distributions are totally different. The local spin currents decay exponentially and eventually become zero within the numerical accuracy when flowing along the longitudinal direction of DNA hairpins, as illustrated in Fig.~\ref{fig3}(e) and the left part of Fig.~\ref{fig3}(b), because $E$ locates outside the electronic bands and all the nucleobases function as strong potential barriers. Subsequently, the electrons passing through DNA hairpins are mainly contributed by the first base pair [Figs.~\ref{fig3}(b) and \ref{fig3}(e)], and the quantum interference effect is negligible. As a result, the oscillating behavior of $G_\uparrow$ vs $E$ cannot be observed in the smooth regions [Figs.~\ref{fig2}(c) and \ref{fig2}(d)]. Besides, one notices that the local spin-up and spin-down currents possess identical direction and will not exhibit vortex-like patterns. In contrast, for model I the local spin currents reappear in the right part of DNA hairpins and are gradually enhanced when the nucleobases are farther away from the source/drain [see the right part of Fig.~\ref{fig3}(b)], due to the finite voltage of lead 3. The local spin currents flow in opposite directions accompanied by a series of vortices. However, the spin polarization is exactly zero in the smooth regions as the source-drain current is hardly affected by lead 3.

\begin{figure}
\includegraphics[width=8.3cm]{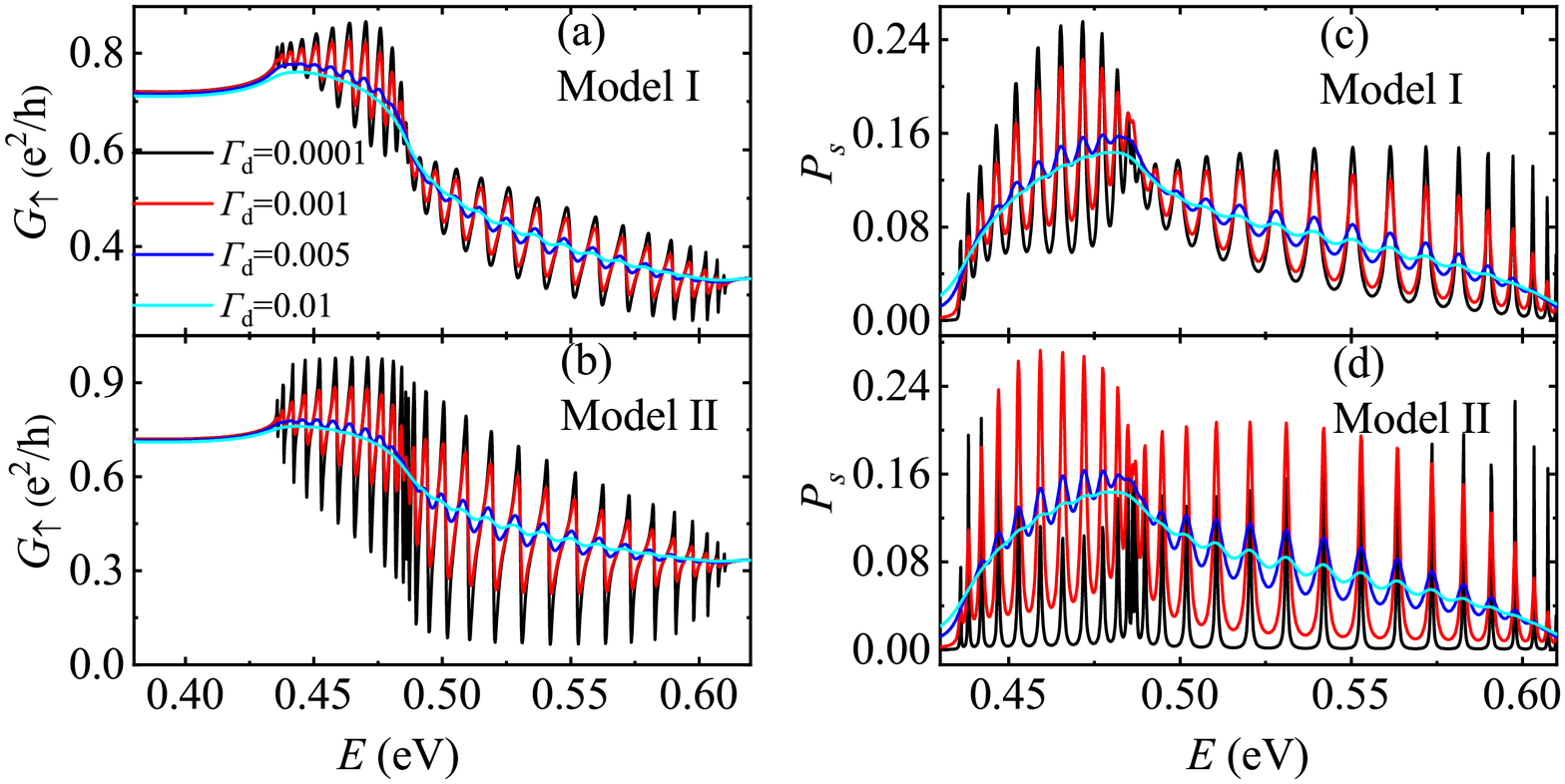}
\caption{\label{fig4} Spin transport along DNA hairpins with a variety of dephasing strengths. $G_{\uparrow }$ vs $E$ for (a) model I and (b) model II; $P_{s}$ vs $E$ for (c) model I and (d) model II. The different lines in each panel denote different dephasing strengths $\Gamma_d $.}
\end{figure}

As the transmission spectra of DNA hairpins are similar between the left energy range $E < E_0$ and the right one $E > E_0$, we focus on the right energy range for clarity. Figures~\ref{fig4}(a)-\ref{fig4}(b) and \ref{fig4}(c)-\ref{fig4}(d) plot $G_\uparrow$ vs $E$ and $P_s$ vs $E$, respectively, for both models with typical values of $\Gamma_d$. It clearly appears that the curves $G_\uparrow$-$E$ and $P_s$-$E$ oscillate around a certain one for different $\Gamma_d$'s. The oscillating amplitude of $G_\uparrow/P_s$ is large in the weak dephasing regime (see the black lines in Fig.~\ref{fig4}) and decreases with increasing $\Gamma_d$ for both models as expected, because the electrons suffer stronger inelastic scattering from the nucleobases in the regime of larger $\Gamma_d$ and the quantum interference effect becomes weaker. Although the oscillating amplitude of $G_\uparrow/P_s$ is smaller in model I than that in model II, their difference becomes smaller with increasing $\Gamma_d$, as the electrons lose their phase memory more quickly in the regime of larger $\Gamma_d$ and then the phase memory loss induced by lead 3 will be reduced. In other words, the hairpin loop has little effect on the spin transport along DNA hairpins in the strong dephasing regime.

The transmission ability of DNA hairpins, however, is very robust against the dephasing. $G_\uparrow$ is slightly declined in the smooth regions and can be enhanced around the valley regions even if $\Gamma_d$ is increased by two orders of magnitude, in sharp contrast to the dsDNA molecules whose conductance decreases with $\Gamma_d$ \cite{gam1}. Further studies indicate that the averaged conductance $\langle G_\uparrow \rangle$, obtained from the right oscillation region, increases with $\Gamma_d$ for $\Gamma_d< \Gamma_d^g$ and decreases with $\Gamma_d$ for $\Gamma_d> \Gamma_d^g$ [Figs.~\ref{fig9}(a) and \ref{fig9}(b)]. Here, $\Gamma_d^g \simeq 0.002$ for model I, which is smaller than the critical value $\Gamma_d^g \simeq 0.004$ of model II, another signature that lead 3 indeed generates additional dephasing in the former model. This dephasing-assisted electron transport in the weak dephasing regime originates from the fact that with increasing $\Gamma_d$, the electrons will preferentially pass through DNA hairpins via the base pairs neighboring to the source/drain and not suffer inelastic scattering from distant nucleobases, leading to the enhancement of the conductance.

More importantly, the spin-filtering effect of DNA hairpins is pronounced for different values of $\Gamma_d$. $P_s$ is nonzero within the entire oscillation regions and can also be enhanced around the valley regions. When $\Gamma_d =0.0001$, $P_s$ can be 25.5\% (22.6\%) for model I (II). When $\Gamma_d =0.01$, $P_s$ can still be 14.4\% for both models. We then calculate the averaged spin polarization, $\langle P_s \rangle$, which is defined as:
\begin{equation}
\langle P_s \rangle =\frac {\langle G_\uparrow \rangle -\langle G_\downarrow \rangle} {\langle G_\uparrow \rangle +\langle G_\downarrow \rangle }.
\end{equation}
Our results indicate that the dependence of $\langle P_s \rangle$ on $\Gamma_d$ is nonmonotonic as well, where $\langle P_s \rangle$ increases with $\Gamma_d$ for $\Gamma_d< \Gamma_d^p$ and decreases with $\Gamma_d$ for $\Gamma_d> \Gamma_d^p$ [Figs.~\ref{fig9}(c) and \ref{fig9}(d)]. Here, $\Gamma_d^p \simeq 0.0035$ (0.0055) for model I (II), which is slightly larger than $\Gamma_d^g$ of model I (II). This nonmonotonic behavior arises from the competing effects of the openness and the phase memory loss induced by the connection to B\"{u}ttiker's virtual leads, where the former effect can generate the spin selectivity and conversely the latter one diminishes the spin polarization \cite{gam1,gam2}. In the weak dephasing regime, the system becomes more open for larger $\Gamma_d$ and leads to higher $\langle P_s \rangle$. While in the strong dephasing regime, the electrons have already lost their phase memory considerably and the spin polarization will be declined by further increasing $\Gamma_d$.

\subsection{\label{sec3B}Effect of molecular length on spin transport along DNA hairpins}

\begin{figure}
\includegraphics[width=8.3cm]{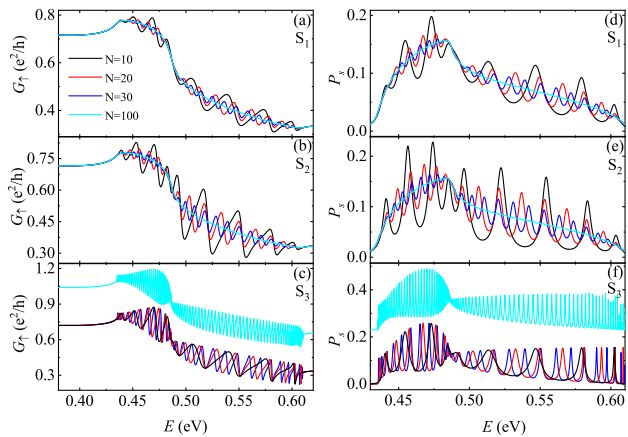}
\caption{\label{fig5} Spin transport along DNA hairpins with a variety of molecular lengths. $G_{\uparrow }$ vs $E$ for (a) $S_1$, (b) $S_2$, and (c) $S_3$; $P_{s}$ vs $E$ for (d) $S_1$, (e) $S_2$, and (f) $S_3$. Here, $S_1$ ($S_2$) refers to model I (model II) with $\Gamma_d =0.005$, and $S_3$ to model I with $\Gamma_d =0$. The different lines denote different molecular lengths $N$. The cyan lines in (c) and (f) are offset for clarity. }
\end{figure}

As $P_s=0$ always holds for model II in the absence of dephasing, we will consider three cases in the following:

($S_{1}$) Model I with $\Gamma_{d}=0.005$,

($S_{2}$) Model II with $\Gamma_{d}=0.005$, and

($S_{3}$) Model I with $\Gamma_{d}=0$.

We then study the effect of the molecular length $N$ on the spin transport along DNA hairpins. Figures~\ref{fig5}(a)-\ref{fig5}(c) and \ref{fig5}(d)-\ref{fig5}(f) show $G_\uparrow$ vs $E$ and $P_s$ vs $E$, respectively, for $S_i$ ($i=1,2,3$) with several values of $N$. One can see that in the presence of B\"{u}ttiker's virtual leads, the curves $G_\uparrow$-$E$ and $P_s$-$E$ of both models also oscillate around a certain one for all investigated values of $N$ [Figs.~\ref{fig5}(a), \ref{fig5}(b), \ref{fig5}(d), and \ref{fig5}(e)]. The oscillating amplitude of $G_\uparrow/P_s$ is large for short DNA hairpins and declined by increasing $N$, because the electrons will suffer inelastic scattering from more and more nucleobases when transmitting through longer DNA hairpins in the case of $\Gamma_d \neq0$ and consequently the quantum interference effect is weakened. While in the absence of B\"{u}ttiker's virtual leads, $G_\uparrow/P_s$ oscillates between two certain envelopes and the corresponding oscillating amplitude remains unchanged for different $N$'s [Figs.~\ref{fig5}(c) and \ref{fig5}(f)], because the inelastic scattering that the electrons suffer only comes from lead 3 which is independent of $N$. Since the number of the electronic states is proportional to $N$, the oscillation frequency of $G_\uparrow/P_s$ increases with $N$.

\begin{figure}
\includegraphics[width=8.3cm]{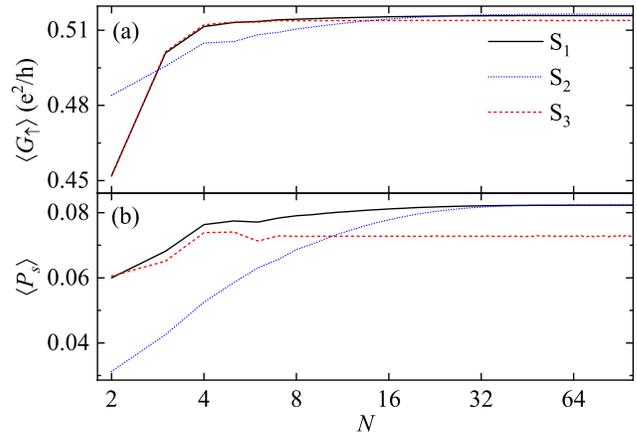}
\caption{\label{fig6} Length-dependent spin transport along DNA hairpins. (a) Averaged spin-up conductance $\left\langle G_{\uparrow}\right\rangle$ vs molecular length $N$ and (b) averaged spin polarization $\left\langle P_{s}\right\rangle$ vs $N$ for $S_1$ (black-solid lines), $S_{2}$ (cyan-dotted lines), and $S_{3}$ (red-dashed lines).}
\end{figure}

Contrary to the physical intuition that the transmission ability should be poorer for longer molecular wires, just as in the dsDNA molecules \cite{gam1}, it is interesting that the averaged conductance, $\langle G_\uparrow \rangle$, could be enhanced by increasing $N$ for both models, as illustrated in Fig.~\ref{fig6}(a). This arises from the increasing probability of electron transmission through DNA hairpins mediated by the base pairs close to the source/drain, similar to the aforementioned dephasing-assisted electron transport which can be further confirmed by the following phenomena. For model I, $\langle G_\uparrow \rangle$ is larger in the case of $\Gamma_d=0.005$ than the case of $\Gamma_d= 0$ when the molecular length satisfies $N>6$ [see the black-solid and red-dashed lines in Fig.~\ref{fig6}(a)]. Although additional dephasing is introduced by lead 3 in model I, its $\langle G_\uparrow \rangle$ is greater than that of model II for relatively short DNA hairpins with $N= 3\sim20$ [see the black-solid and blue-dotted lines in Fig.~\ref{fig6}(a)]. This implies that the transmission ability of short DNA hairpins is sensitive to their hairpin loops, in agreement with previous experiments \cite{lfd3,rn2}. While in the smooth regions, $G_\uparrow$ is almost independent of $N$ for all $S_i$'s as expected [Figs.~\ref{fig5}(a)-\ref{fig5}(c)], because the electron transmission in these areas is dominated by the first base pair and will not be affected by distant nucleobases.

By inspecting Fig.~\ref{fig6}(b), one may notice that the averaged spin polarization depends on the dephasing strength and the hairpin loop. This phenomenon is related to the openness induced by the connection to extra real/virtual leads. On the one hand, since the number of B\"{u}ttiker's virtual leads is proportional to $N$, the system becomes more open for longer DNA hairpins when $\Gamma_d \neq0$. As a result, model I with $\Gamma_d=0.005$ exhibits higher $\langle P_s \rangle$ as compared with the case of $\Gamma_d=0$, and their difference increases with $N$ [see the black-solid and red-dashed lines in Fig.~\ref{fig6}(b)]. On the other hand, since lead 3 promotes the openness of model I, its $\langle P_s \rangle$ can be greater than that of model II for short DNA hairpins with $N<11$ [Fig.~\ref{fig6}(b)], indicating that lead 3 plays an important role in the spin transport along short DNA hairpins. Nevertheless, the difference of $\langle P_s\rangle$ between both models is decreased by increasing $N$ [see the black-solid and blue-dotted lines in Fig.~\ref{fig6}(b)], because the system cannot feel additional openness coming from lead 3 when it is sufficient open for long DNA hairpins. This implies that the spin transport properties are almost independent of the hairpin loop when the molecular length is sufficiently long.

Furthermore, one can see that $\langle P_s \rangle$ increases with $N$ for short DNA hairpins and then saturates at a critical molecular length $N_c$ for all $S_i$'s. The critical molecular length depends strongly on the dephasing strength and the hairpin loop, where $N_c=21$, $31$, and $5$ for $S_1$, $S_2$, and $S_3$, respectively. This behavior arises from the competition between the openness and the phase memory loss, which compensate for each other in the regime of large $N$ and thus $\langle P_s \rangle$ saturates for long DNA hairpins. Despite their strong dependence upon the dephasing and the loop, DNA hairpins exhibit pronounced spin-filtering effect for different molecular lengths. When $N=10$, $P_s$ can be 19.8\% (22.8\%) and $\langle P_s \rangle\simeq 8.0\%$ (7.2\%) for model I (II); when $N=100$, $P_s$ can still be 15.6\% and $ \langle P_s \rangle\simeq 8.2\%$ for both models.

\subsection{\label{sec3C}Effect of interchain coupling on spin transport along DNA hairpins}

\begin{figure}
\includegraphics[width=8.3cm]{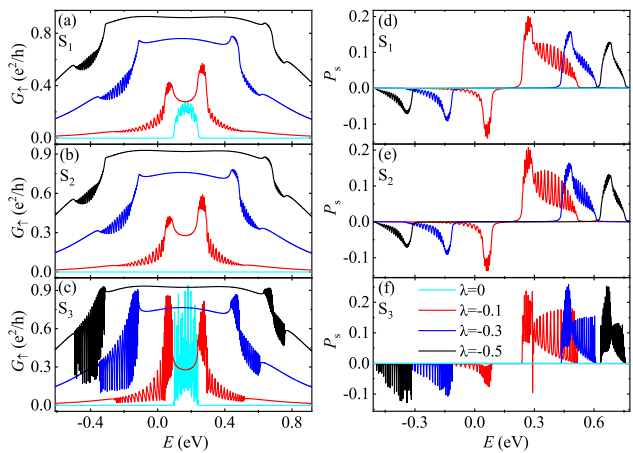}
\caption{\label{fig7} Spin transport along DNA hairpins with several values of the interchain couplings. $G_{\uparrow }$ vs $E$ for (a) $S_1$, (b) $S_2$, and (c) $S_3$; $P_s$ vs $E$ for (d) $S_1$, (e) $S_2$, and (f) $S_3$. The different lines represent various interchain couplings $\lambda$.}
\end{figure}

We now turn to study the influence of the interchain coupling $\lambda$ on the spin transport along DNA hairpins. Figures~\ref{fig7}(a)-\ref{fig7}(c) and \ref{fig7}(d)-\ref{fig7}(f) show $G_\uparrow$ vs $E$ and $P_s$ vs $E$, respectively, with $\lambda$ changing from $0$ to $-0.5$. When the interchain coupling is switched off, the two helical chains decouple with each other. Then, the electrons cannot transmit within the base pairs and the transport properties are mainly determined by the hairpin loop. For model I where the hairpin loop is conductive, the electrons can transport from the source to the drain mediated by the conducting loop and a transmission band, associated with the electronic structures of both helical chains, emerges around the band center [see the cyan lines in Figs.~\ref{fig7}(a) and \ref{fig7}(c)]. In this case, however, the spin-filtering effect disappears for whatever the values of model parameters [see the cyan lines in Figs.~\ref{fig7}(d) and \ref{fig7}(f)], because the system is simplified as a single-helical chain with only one transport pathway, consistent with previous works \cite{gb1,gam1,gam2}. While for model II where the hairpin loop is insulating, there does not exist any transport pathway. Then, both $G_\uparrow$ and $G_\downarrow$ are exactly zero [see the cyan line in Fig.~\ref{fig7}(b)], and $P_s$ is uncertain.

When the interchain coupling is switched on, the original transmission band at $\lambda=0$ evolves into two oscillation regions which are separated by a smooth one, exhibiting distinct features in comparison to the case of $\lambda=0$. With increasing $\lambda$, one can see several general phenomena for both models. (i) $G_\uparrow$ is considerably increased over the entire energy spectrum, because the electron transmission is enhanced along the transverse direction and suppressed along the longitudinal one. In other words, the probability of electron transmission from the source to the drain, mediated by the neighboring base pairs and not suffering any scattering from the distant nucleobases, is strongly increased, which is analogous to the electron transport mechanism found in the smooth regions. Subsequently, the quantum interference effect becomes weaker and the oscillating amplitude of $G_\uparrow/P_s$ decreases in general, accompanied by the decrement of $P_s$ [Figs.~\ref{fig7}(d) and \ref{fig7}(e)]. (ii) The two oscillation regions are further separated from each other and become narrower, owing to the repulsion effect between the two helical chains driven by the interchain coupling, so do the energy regions with finite $P_s$ [Figs.~\ref{fig7}(d)-\ref{fig7}(f)]. In sharp contrast, the width of the smooth regions is dramatically increased, as can be seen from, e.g., the central smooth regions in Figs.~\ref{fig7}(a)-\ref{fig7}(c). As a result, the energy spectrum of finite $G_\uparrow$ becomes wider with increasing $\lambda$. (iii) More importantly, although $P_s$ is reduced by increasing $\lambda$ in the presence of B\"{u}ttiker's virtual leads, the spin-filtering effect remains significant for various $\lambda$'s, especially in the case of $\Gamma_d=0$ [Fig.~\ref{fig7}(f)]. For example, when $\lambda=-0.5$, $P_s$ can be 13.2\% for model II with $\Gamma_d=0.005$ and 25.3\% for model I with $\Gamma_d=0$.

\subsection{\label{sec3D}Contact effect on spin transport along DNA hairpins}

\begin{figure}
\includegraphics[width=8.3cm]{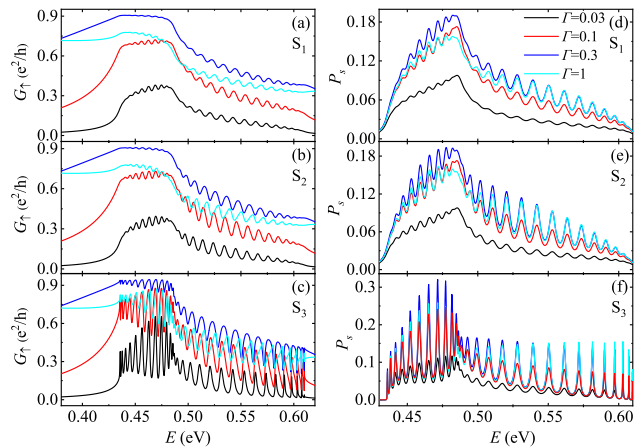}
\caption{\label{fig8} Spin transport along DNA hairpins with several values of the molecule-lead couplings. $G_{\uparrow }$ vs $E$ for (a) $S_1$, (b) $S_2$, and (c) $S_3$; $P_s$ vs $E$ for (d) $S_1$, (e) $S_2$, and (f) $S_3$. The legends in (d) are for all panels.}
\end{figure}

\begin{figure}
\subfigure{ \label{fig:subfig:a} 
\includegraphics[width=4.1cm]{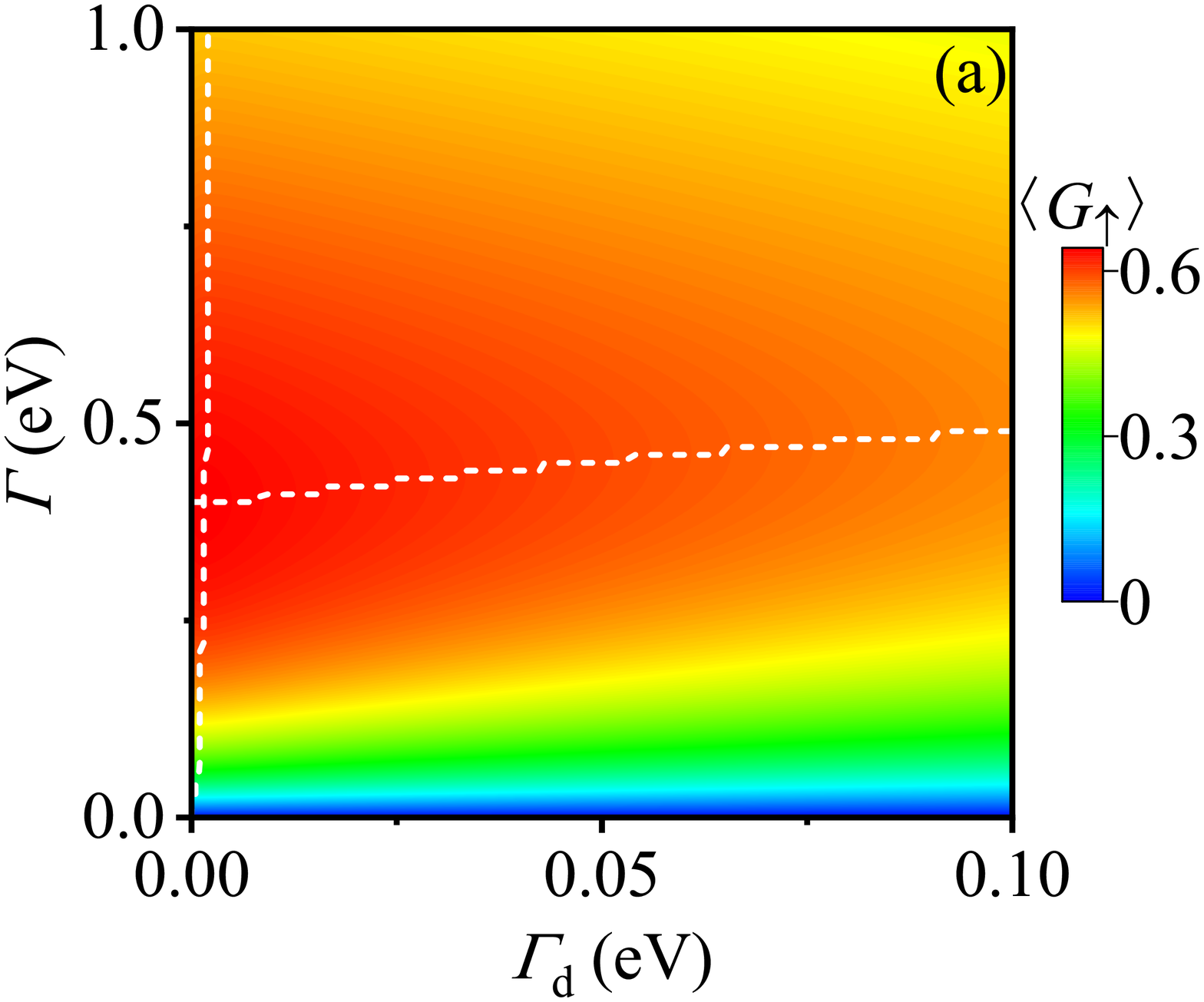}}
\subfigure{ \label{fig:subfig:b} 
\includegraphics[width=4.1cm]{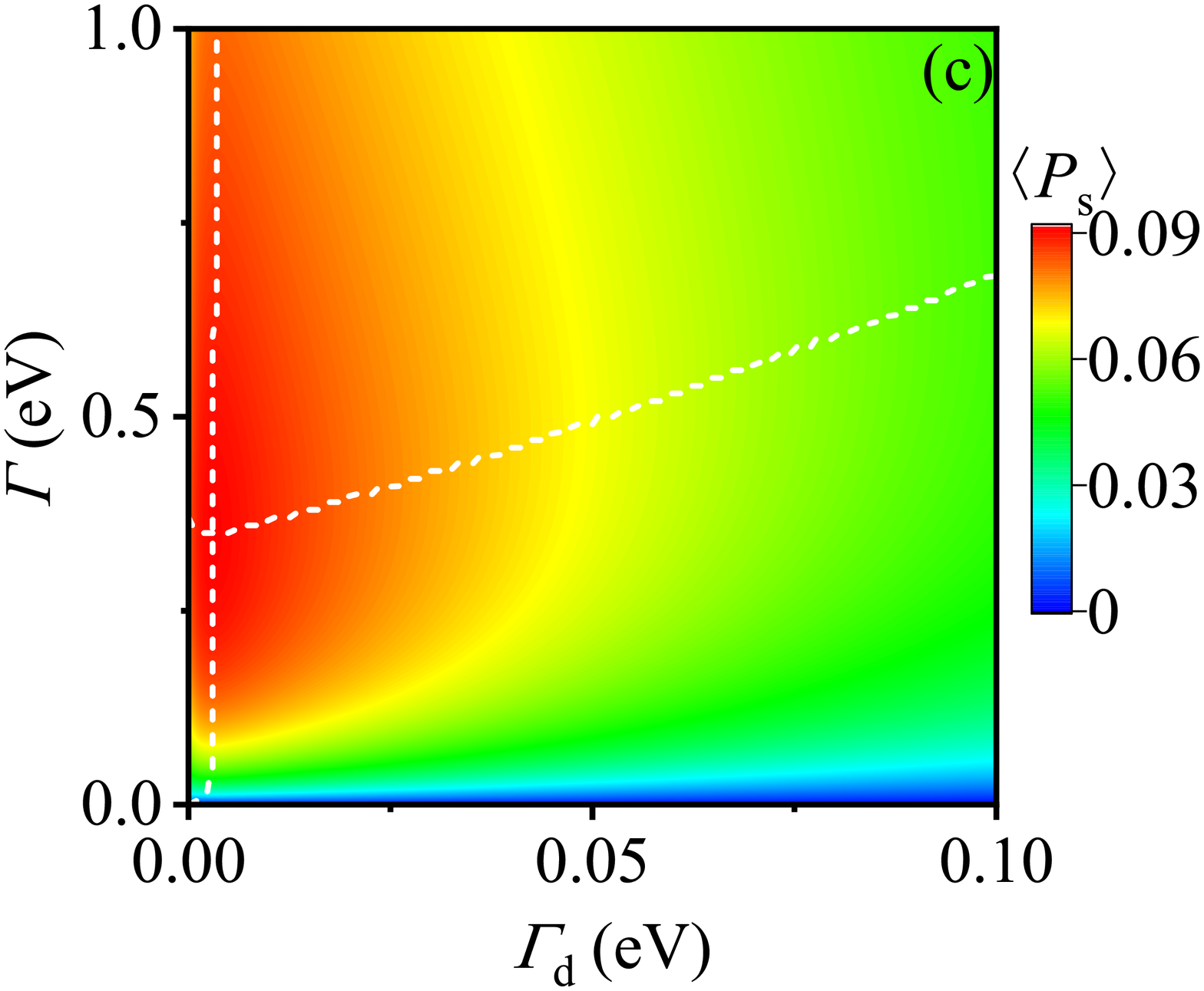}}
\subfigure{ \label{fig:subfig:c} 
\includegraphics[width=4.1cm]{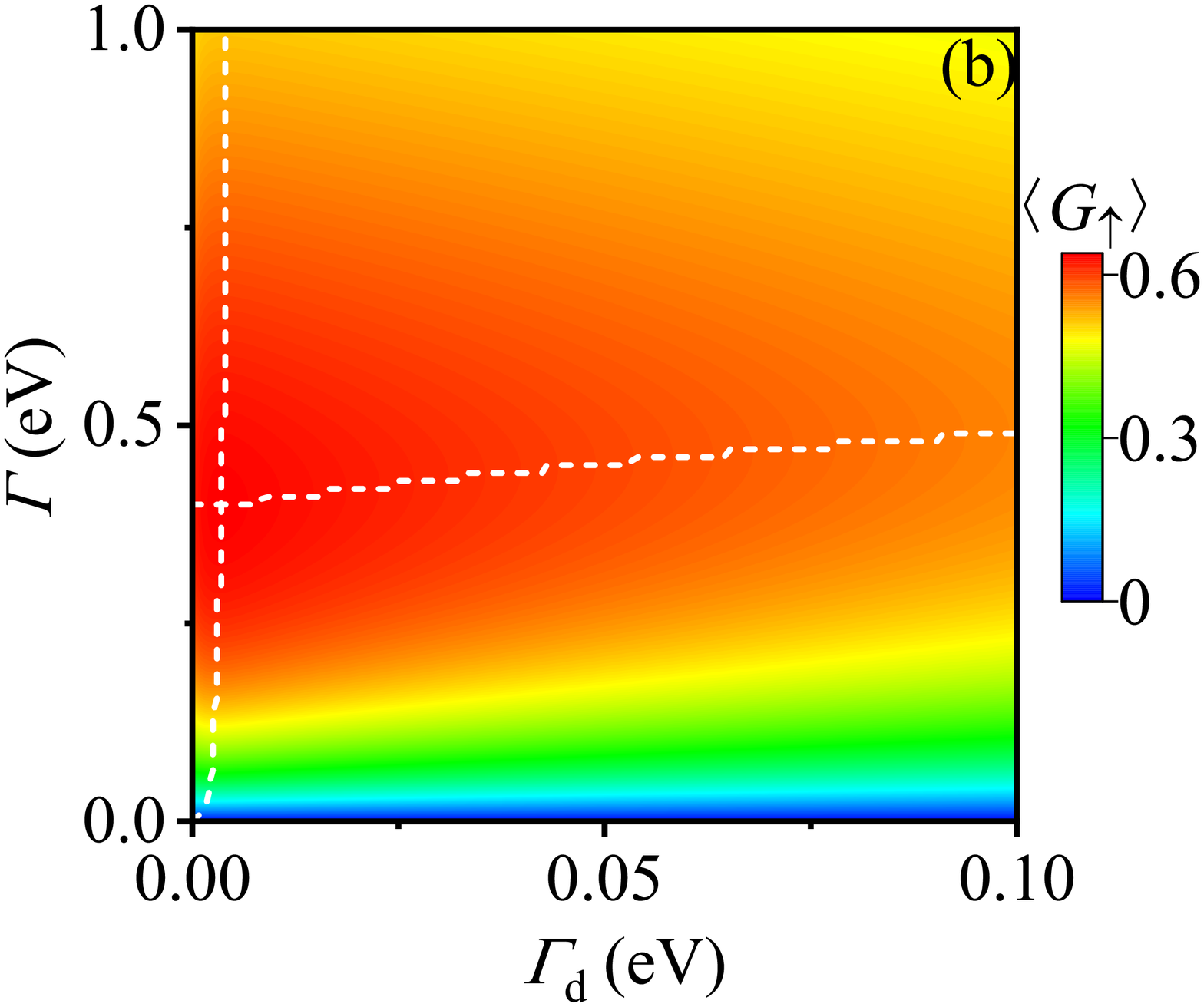}}
\subfigure{ \label{fig:subfig:d} 
\includegraphics[width=4.1cm]{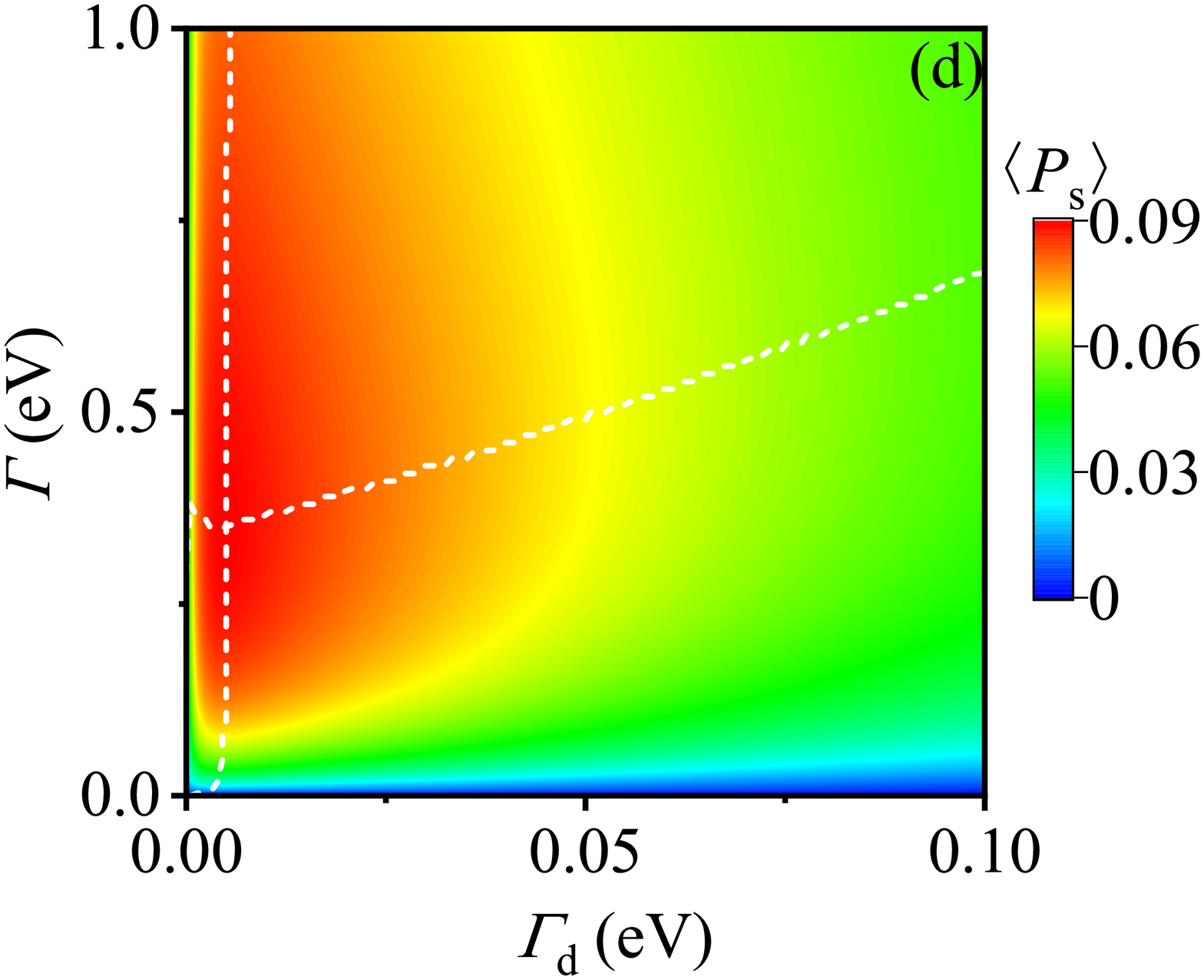}}
\caption{Spin transport along DNA hairpins by taking into account various dephasing strengths and molecule-lead couplings. 2D plot of $\left\langle G_{\uparrow }\right\rangle$ vs $\Gamma_d$ and $\Gamma$ for (a) model I and (b) model II; 2D plot of $\left\langle P_{s}\right\rangle$ vs $\Gamma_d$ and $\Gamma$ for (c) model I and (d) model II. In (a) and (b), the horizontal and vertical white-dashed lines denote, respectively, the turning points $\Gamma^g$ in the curves $\langle G_\uparrow \rangle$-$\Gamma$ and $\Gamma_d^g$ in the curves $\langle G_\uparrow \rangle$-$\Gamma_d$. While in (c) and (d), the oblique and vertical white-dashed lines represent, respectively, the turning points $\Gamma^p$ in the curves $\langle P_s \rangle$-$\Gamma$ and $\Gamma_d^p$ in the curves $\langle P_s \rangle$-$\Gamma_d$.}
\label{fig9} 
\end{figure}

Finally, we investigate the contact effect on the spin transport along DNA hairpins. The contact between the molecule and the leads in various experiments can generally be divided into two categories, i.e., the physisorbed contact and the chemisorbed one. In the former case, the end sites of the molecule are directly attracted to the leads by electrostatic trapping and the contact is usually poor. In the latter case, the contact is realized by chemical bonding between the end sites and the leads via thiol groups, allowing reproducible transport measurements. The contact effect can then be simulated by considering different coupling strengths between the hairpin stem and the source/drain. Figures~\ref{fig8}(a)-\ref{fig8}(c) and \ref{fig8}(d)-\ref{fig8}(f) show $G_\uparrow$ vs $E$ and $P_s$ vs $E$, respectively, for typical values of $\Gamma$. When $\Gamma$ is small, the contact functions as a strong tunnelling barrier and majority of the electrons will be reflected at the interface between the source and the hairpin stem. As a result, $G_\uparrow$ is small in the oscillation regions and becomes smaller in the smooth ones due to the stronger suppression of the electron transmission along the first base pair [see the black lines in Figs.~\ref{fig8}(a)-\ref{fig8}(c)]. By increasing $\Gamma$, the electron transmission across the interface is enhanced and the conductance is increased within most of the smooth regions (data not shown). Nevertheless, the conductance profiles in the oscillation regions are distinct and present nonmonotonic behavior, which can be further demonstrated in Figs.~\ref{fig9}(a) and \ref{fig9}(b) where the two-dimensional (2D) plots of $\langle G_\uparrow \rangle$ for both models are displayed as functions of $\Gamma_d$ and $\Gamma$.

One can see from Figs.~\ref{fig9}(a) and \ref{fig9}(b) that by fixing $\Gamma_d$, a turning point can always be observed in the curves $\langle G_\uparrow \rangle$-$\Gamma$, where $\langle G_\uparrow \rangle$ increases with $\Gamma$ for $\Gamma < \Gamma ^g$ and decreases with $\Gamma$ for $\Gamma > \Gamma ^g$, regardless of $\Gamma_d$. This turning point increases almost linearly with $\Gamma_d$, from $\Gamma ^g \simeq0.4$ at $\Gamma_d=0$ to $\Gamma ^g \simeq0.49$ at $\Gamma_d=0.1$ for both models, as illustrated by the horizontal white-dashed lines in Figs.~\ref{fig9}(a) and \ref{fig9}(b).  This optimal contact configuration for efficient electron transport through DNA hairpins is different from other one-dimensional molecular wires \cite{zy1,gam4} and can be controlled by the dephasing. While fixing $\Gamma$, there also exists a turning point in the curves $\langle G_\uparrow \rangle$-$\Gamma_d$ that $\langle G_\uparrow \rangle$ increases with $\Gamma_d$ for $\Gamma_d <\Gamma_d^g$ and decreases with $\Gamma_d$ for $\Gamma_d >\Gamma_d^g$, as can be seen from the vertical white-dashed lines in Figs.~\ref{fig9}(a) and \ref{fig9}(b). This indicates that the dephasing-assisted electron transport is a general phenomenon for DNA hairpins in the weak dephasing regime. One may notice that $\Gamma_d^g$ of model I is always smaller than that of model II as lead 3 introduces additional dephasing in the former model I, and both $\Gamma_d^g$'s increase slowly with $\Gamma$. Besides, although the electrons are strongly reflected at the interface in the regime of weak $\Gamma$, DNA hairpins exhibit strong transmission ability in a very wide range of $\Gamma_d$ and $\Gamma$ with $\langle G_\uparrow \rangle>0.1e^2/h$, as illustrated by the colorful areas except the blue ones in Figs.~\ref{fig9}(a) and \ref{fig9}(b).

By inspecting Figs.~\ref{fig8}(d)-\ref{fig8}(f), it clearly appears that the spin polarization increases with $\Gamma$ at first and is then suppressed by further increasing $\Gamma$, accompanied by the increment of the oscillating amplitude of $P_s$. This nonmonotonic behavior can also be detected in Figs.~\ref{fig9}(c) and \ref{fig9}(d), where the 2D plots of $\langle P_s \rangle$ vs $\Gamma_d$ and $\Gamma$ are shown for both models. When $\Gamma_d$ is fixed, $\langle P_s \rangle$ increases with $\Gamma$ for $\Gamma< \Gamma ^p$ and decreases with $\Gamma$ for $\Gamma> \Gamma ^p$, for whatever the values of $\Gamma_d$, as illustrated by the oblique white-dashed lines in Figs.~\ref{fig9}(c) and \ref{fig9}(d). Interestingly, this optimized contact condition of the spin-selectivity effect is distinct from that of efficient electron transport mentioned above, $\Gamma^p \neq \Gamma^g$. Here, $\Gamma^p$ is decreased from $\Gamma ^p \simeq 0.36$ (0.38) at $\Gamma_d=0.0005$ to $\Gamma ^p \simeq 0.35$ at $\Gamma_d=0.001$ (0.003) and then increased almost linearly to $\Gamma ^p \simeq 0.69$ at $\Gamma_d=0.1$ for model I (II). When $\Gamma$ is fixed, the dependence of $\langle P_s \rangle$ on $\Gamma_d$ is also nonmonotonic that $\langle P_s \rangle$ increases with $\Gamma_d$ for $\Gamma< \Gamma_d^p$ and decreases with $\Gamma_d$ for $\Gamma> \Gamma_d^p$, regardless of $\Gamma$. This turning point depends weakly on $\Gamma$ and will be slightly increased from $\Gamma_d ^p \simeq 0.002$ (0.0035) at $\Gamma=0.01$ to $\Gamma_d ^p \simeq 0.0035$ (0.0055) at $\Gamma=1$ for model I (II) [see the vertical white-dashed lines in Figs.~\ref{fig9}(c) and \ref{fig9}(d)]. Note that $\Gamma_d^p > \Gamma_d^g$ always for either model of DNA hairpins.

Besides, we find that $\langle P_s \rangle$ of model I is greater than that of model II in the weak dephasing regime [see the leftmost part of Figs.~\ref{fig9}(c) and \ref{fig9}(d)], as lead 3 promotes the openness of the former model and plays a dominant role in this regime. While in the strong dephasing regime, there is no observable difference of $\langle P_s \rangle$ between both models, further demonstrating that the hairpin loop will hardly influence the spin transport properties of DNA hairpins. Despite the complicated dependence of the spin polarization on the dephasing strength and the molecule-lead coupling, the spin-filtering effect of DNA hairpins is pronounced in a wide range of $\Gamma_d$ and $\Gamma$. For example, when $\Gamma_d=0.002$ and $\Gamma=0.2$, $P_s$ can reach 22.8\% (26.7\%) and $\langle P_s \rangle \simeq 8.8\%$ (8.1\%) for model I (II); when $\Gamma_d=0.05$ and $\Gamma=0.8$, $P_s$ can still be 9.6\% and $\langle P_s \rangle \simeq 6.7\%$ for both models.

\section{\label{sec4}CONCLUSIONS}

In summary, the spin transport along DNA hairpins with either conducting loops or insulating ones has been studied, where two nonmagnetic leads are contacted, respectively, at the two ends of the hairpin stem. Using the nonequilibrium Green's function, the conductance and the spin polarization are calculated by considering dephasing strength, molecular length, interchain coupling, and molecule-lead coupling. We find that DNA hairpins present pronounced spin-filtering effect in a wide range of model parameters. As compared with the insulating loops, the conducting loops simulated by a nonmagnetic lead could enhance the spin-filtering effect in the regimes of weak dephasing and of short molecular length. In particular, their spin transport properties exhibit some unique features which are different from double-stranded DNA molecules. (i) When the electron energy locates in the electronic bands of the hairpin stem, the electrons can propagate along the longitudinal direction of DNA hairpins, giving rise to a series of vortex clusters with identical chirality in either oscillation region. Besides, the chirality of the vortex clusters will be reversed by tuning the electron energy between the two oscillation regions, leading to the sign reversal of the spin polarization. The local spin currents can be greater than the corresponding spin component of the source-drain current. When the electron energy locates beyond the electronic bands, the electron transmission decays exponentially along the longitudinal direction and the oscillating behavior of the conductance disappears. (ii) Both the conductance and the spin polarization increase with the molecular length and saturate for sufficiently long DNA hairpins. With increasing the interchain coupling, the conductance is considerably enhanced over the entire energy spectrum and the two oscillation regions are further separated from each other. (iii) The dependence of the conductance and the spin polarization on either the dephasing strength or the molecule-lead coupling is similar among each other, all of which exhibit nonmonotonic behavior. Furthermore, we demonstrate the optimized contact configurations of efficient electron transport and of the spin-selectivity effect, revealing that they are distinct from each other and depend strongly on the dephasing strength.

\section*{Acknowledgments}

This work is supported by the National Natural Science Foundation of China (Grants No. 11874428, No. 11874187, and No. 11921005), the Innovation-Driven Project of Central South University (Grant No. 2018CX044), and the National Key Research and Development Program of China (Grant No. 2017YFA0303301).


\begin{references}

\bibitem{nr1}  R. Naaman and D. H. Waldeck, J. Phys. Chem. Lett. {\bf 3}, 2178 (2012).
\bibitem{ajm1} J. M. Abendroth, D. M. Stemer, B. P. Bloom, P. Roy, R. Naaman, D. H. Waldeck, P. S. Weiss, and P. C. Mondal, ACS Nano {\bf 13}, 4928 (2019).
\bibitem{nr2}  R. Naaman, Y. Paltiel, and D. H. Waldeck, Nat. Rev. Chem. {\bf 3}, 250 (2019).


\bibitem{gb1}  B. G\"{o}hler, V. Hamelbeck, T. Z. Markus, M. Kettner, G. F. Hanne, Z. Vager, R. Naaman, and H. Zacharias, Science {\bf 331}, 894 (2011).
\bibitem{xz1}  Z. Xie, T. Z. Markus, S. R. Cohen, Z. Vager, R. Gutierrez, and R. Naaman, Nano Lett. {\bf 11}, 4652 (2011).
\bibitem{md1}  D. Mishra, T. Z. Markus, R. Naaman, M. Kettner, B. Gohler, H. Zacharias, N. Friedman, M. Sheves, and C. Fontanesi, Proc. Natl. Acad. Sci. (USA) {\bf 110}, 14872 (2013).
\bibitem{dob1} O. B. Dor, S. Yochelis, S. P. Mathew, R. Naaman, and Y. Paltiel, Nat. Commun. {\bf 4}, 2256 (2013).


\bibitem{ztj1} T. J. Zwang, S. H\"{u}rlimann, M. G. Hill, and J. K. Barton, J. Am. Chem. Soc. {\bf 138}, 15551 (2016).
\bibitem{kv1}  V. Kiran, S. P. Mathew, S. R. Cohen, I. H. Delgado, J. Lacour, and R. Naaman, Adv. Mater. {\bf 28}, 1957 (2016).
\bibitem{ajm2} J. M. Abendroth, N. Nakatsuka, M. Ye, D. Kim, E. E. Fullerton, A. M. Andrews, and P. S. Weiss, ACS Nano {\bf 11}, 7516 (2017).
\bibitem{dob2} O. B. Dor, S. Yochelis, A. Radko, K. Vankayala, E. Capua, A. Capua, S.-H. Yang, L. T. Baczewski, S. S. P. Parkin, R. Naaman, and Y. Paltiel, Nat. Commun. {\bf 8}, 14567 (2017).
\bibitem{akm1} K. M. Alam and S. Pramanik, Nanoscale {\bf 9}, 5155 (2017).
\bibitem{km1}  M. Kettner, V. V. Maslyuk, D. N\"{u}renberg, J. Seibel, R. Gutierrez, G. Cuniberti, K.-H. Ernst, and H. Zacharias, J. Phys. Chem. Lett. {\bf 9}, 2025 (2018).
\bibitem{ms1}  S. Mishra, V. S. Poonia, C. Fontanesi, R. Naaman, A. M. Fleming, and C. J. Burrows, J. Am. Chem. Soc. {\bf 141}, 123 (2019).
\bibitem{ajm3} J. M. Abendroth, K. M. Cheung, D. M. Stemer, M. S. El Hadri, C. Zhao, E. E. Fullerton, and P. S. Weiss, J. Am. Chem. Soc. {\bf 141}, 3863 (2019).
\bibitem{sm1}  M. Suda, Y. Thathong, V. Promarak, H. Kojima, M. Nakamura, T. Shiraogawa, M. Ehara, and H. M. Yamamoto, Nat. Commun. {\bf 10}, 2455 (2019).


\bibitem{gam1} A.-M. Guo and Q.-F. Sun, Phys. Rev. Lett. {\bf 108}, 218102 (2012).
\bibitem{gam2} A.-M. Guo and Q.-F. Sun, Proc. Natl. Acad. Sci. (USA) {\bf 111}, 11658 (2014).
\bibitem{me1}  E. Medina, F. L\'{o}pez, M. A. Ratner, and V. Mujica, Europhys. Lett. {\bf 99}, 17006 (2012).
\bibitem{eaa1} A. A. Eremko and V. M. Loktev, Phys. Rev. B {\bf 88}, 165409 (2013).
\bibitem{gam3} A.-M. Guo, E. D\'{i}az, C. Gaul, R. Gutierrez, F. Dom\'{i}nguez-Adame, G. Cuniberti, and Q.-F. Sun, Phys. Rev. B {\bf 89}, 205434 (2014).
\bibitem{vs1}  S. Varela, V. Mujica, and E. Medina, Phys. Rev. B {\bf 93}, 155436 (2016).
\bibitem{whn1} H.-N. Wu, X. Wang, Y.-J. Zhang, G.-Y. Yi, and W.-J. Gong, Appl. Phys. A {\bf 122}, 626 (2016).
\bibitem{mas1} S. Matityahu, Y. Utsumi, A. Aharony, O. Entin-Wohlman, and C. A. Balseiro, Phys. Rev. B {\bf 93}, 075407 (2016).
\bibitem{ay1}  Y. Avishai, and Y. B. Band, Phys. Rev. B {\bf 95}, 104429 (2017).
\bibitem{de1}  E. D\'{i}az, A. Contreras, J. Hern\'{a}ndez, and F. Dom\'{i}nguez-Adame, Phys. Rev. E {\bf 98}, 052221 (2018).
\bibitem{mvv1} V. V. Maslyuk, R. Gutierrez, A. Dianat, V. Mujica, and G. Cuniberti, J. Phys. Chem. Lett. {\bf 9}, 5453 (2018).
\bibitem{mk1}  K. Michaeli and R. Naaman, J. Phys. Chem. C {\bf 123}, 17043 (2019).
\bibitem{ds1}  S. Dalum and P. Hedeg{\aa}rd, Nano Lett. {\bf 19}, 5253 (2019).
\bibitem{nd1}  D. N\"{u}renberg and H. Zacharias, Phys. Chem. Chem. Phys. {\bf 21}, 3761 (2019).
\bibitem{gm1}  M. Geyer, R. Gutierrez, V. Mujica, and G. Cuniberti, J. Phys. Chem. C {\bf 123}, 27230 (2019).
\bibitem{yx1}  X. Yang, C. H. van der Wal, and B. J. van Wees, Phys. Rev. B {\bf 99}, 024418 (2019).
\bibitem{dgf1} G.-F. Du, H.-H. Fu, and R. Wu, Phys. Rev. B {\bf 102}, 035431 (2020).


\bibitem{lfd1} F. D. Lewis, X. Liu, J. Liu, S. E. Miller, R. T. Hayes, and M. R. Wasielewski, Nature {\bf 406}, 51 (2000).
\bibitem{lfd2} F. D. Lewis, H. Zhu, P. Daublain, T. Fiebig, M. Raytchev, Q. Wang, and V. Shafirovich, J. Am. Chem. Soc. {\bf 128}, 791 (2006).
\bibitem{kk1}  K. Kawai and T. Majima, Acc. Chem. Res. {\bf 46}, 2616 (2013).
\bibitem{lfd3} F. D. Lewis, R. M. Young, and M. R. Wasielewski, Acc. Chem. Res. {\bf 51}, 1746 (2018).


\bibitem{lfd4} F. D. Lewis, T. Wu, Y. Zhang, R. L. Letsinger, S. R. Greenfield, and M. R. Wasielewski, Science {\bf 277}, 673 (1997).
\bibitem{gfc1} F. C. Grozema, S. Tonzani, Y. A. Berlin, G. C. Schatz, L. D. A. Siebbeles, and M. A. Ratner, J. Am. Chem. Soc. {\bf 130}, 5157 (2008).
\bibitem{gfc2} F. C. Grozema, S. Tonzani, Y. A. Berlin, G. C. Schatz, L. D. A. Siebbeles, and M. A. Ratner, J. Am. Chem. Soc. {\bf 131}, 14204 (2009).
\bibitem{rn1}  N. Renaud, Y. A. Berlin, F. D. Lewis, and M. A. Ratner, J. Am. Chem. Soc. {\bf 135}, 3953 (2013).


\bibitem{rn2}  N. Renaud, M. A. Harris, A. P. N. Singh, Y. A. Berlin, M. A. Ratner, M. R. Wasielewski, F. D. Lewis, and F. C. Grozema, Nat. Chem. {\bf 8}, 1015 (2016).
\bibitem{zta1} T. A. Zeidan, R. Carmieli, R. F. Kelley, T. M. Wilson, F. D. Lewis, and M. R. Wasielewski, J. Am. Chem. Soc. {\bf 130}, 13945 (2008).
\bibitem{cr1}  R. Carmieli, A. K. Thazhathveetil, F. D. Lewis, and M. R. Wasielewski, J. Am. Chem. Soc. {\bf 135}, 10970 (2013).
\bibitem{ojh1} J. H. Olshansky, M. D. Krzyaniak, R. M. Young, and M. R. Wasielewski, J. Am. Chem. Soc. {\bf 141}, 2152 (2019).
\bibitem{ojh2} J. H. Olshansky, J. Zhang, M. D. Krzyaniak, E. R. Lorenzo, and M. R. Wasielewski, J. Am. Chem. Soc. {\bf 142}, 3346 (2020).
\bibitem{sdm1} D. M. Stemer, J. M. Abendroth, K. M. Cheung, M. Ye, M. S. El Hadri, E. E. Fullerton, and P. S. Weiss, Nano Lett. {\bf 20}, 1218 (2020).


\bibitem{bg1}  G. Bonnet, O. Krichevsky, and A. Libchaber, Proc. Natl. Acad. Sci. (USA) {\bf 95}, 8602 (1998).
\bibitem{gnl1} N. L. Goddard, G. Bonnet, O. Krichevsky, and A. Libchaber, Phys. Rev. Lett. {\bf 85}, 2400 (2000).
\bibitem{wmi1} M. I. Wallace, L. Ying, S. Balasubramanian, and D. Klenerman, Proc. Natl. Acad. Sci. (USA) {\bf 98}, 5584 (2001).
\bibitem{ts1}  S. Tyagi and F. R. Kramer, Nat. Biotechnol. {\bf 14}, 303 (1996).
\bibitem{ts2}  S. Tyagi, D. P. Bratu, and F. R. Kramer, Nat. Biotechnol. {\bf 16}, 49 (1998).
\bibitem{db1}  B. Dubertret, M. Calame, and A. J. Libchaber, Nat. Biotechnol. {\bf 19}, 365 (2001).


\bibitem{ds1t} {\it Electronic Transport in Mesoscopic Systems}, edited by S. Datta (Cambridge University Press, Cambridge, England, 1995).
\bibitem{jap1} A.-P. Jauho, N. S. Wingreen, and Y. Meir, Phys. Rev. B {\bf 50}, 5528 (1994).
\bibitem{jh1}  H. Jiang, L. Wang, Q.-F. Sun, and X. C. Xie, Phys. Rev. B {\bf 80}, 165316 (2009).
\bibitem{persistC1} E. I. Rashba, Phys. Rev. B {\bf 68}, 241315(R) (2003).
\bibitem{persistC2} Q.-F. Sun, X. C. Xie, and J. Wang, Phys. Rev. Lett. {\bf 98}, 196801 (2007).
\bibitem{ns1}  S. Nakanishi and M. Tsukada, Phys. Rev. Lett. {\bf 87}, 126801 (2001).
\bibitem{zyy1} Y. Zhang, J.-P. Hu, B. A. Bernevig, X. R. Wang, X. C. Xie, and W. M. Liu, Phys. Rev. B {\bf 78}, 155413 (2008).


\bibitem{sqf1} Q.-F. Sun and X. C. Xie, Phys. Rev. B {\bf 71}, 155321 (2005).
\bibitem{sgc1} G. C. Solomon, C. Herrmann, T. Hansen, V. Mujica, and M. A. Ratner, Nat. Chem. {\bf 2}, 223 (2005).


\bibitem{zy1}  Y. Zhu, C.-C. Kaun, and H. Guo, Phys. Rev. B {\bf 69}, 245112 (2004).
\bibitem{gam4} A.-M. Guo and S.-J. Xiong, Phys. Rev. B {\bf 80}, 035115 (2009).

\end{references}
\end{document}